\documentclass[journal=apchd5,manuscript=article]{achemso}
         
\usepackage{amssymb}
\usepackage{amsmath}
\usepackage[font={small}]{caption}

\SectionNumbersOff 
     
\author{F. Le Roux}
\email{florian.leroux@physics.ox.ac.uk}
\affiliation[University of Oxford]
{Department of Physics, University of Oxford, Parks Road, Oxford, OX1 3PU, U.K.}

\author{R. A. Taylor}
\affiliation[University of Oxford]
{Department of Physics, University of Oxford, Parks Road, Oxford, OX1 3PU, U.K.}

\author{D. D. C. Bradley}
\affiliation[University of Oxford]
{Department of Physics, University of Oxford, Parks Road, Oxford, OX1 3PU, U.K.}
\alsoaffiliation[KAUST]
{Physical Science and Engineering Division, King Abdullah University of Science and Technology, Thuwal, 23955-6900, Saudi Arabia}

\title{Angle-resolved Reflectivity and Transmissivity of a TE-polarized Wave Incident On a Microcavity Containing Strongly Coupled In-plane Oriented 1D Excitons}
\begin{document}

\begin{abstract}
Here we report on the reflectivity and transmissivity of a TE-polarized wave incident on a microcavity containing strongly coupled, in-plane oriented one dimensional (1D) excitons. We first discuss the propagation of the electric field through the cavity and present a simple model which allows us to understand the underlying physics. We then compare this model to previous reports and perform our own measurements on a microcavity containing an oriented layer of liquid-crystalline poly(9,9-dioctyfluorene) (PFO). We show that in all cases, the reflected and transmitted electric fields are the superpositions of photons leaking parallel and perpendicular to the excitons' orientation.
\end{abstract}

\newpage

\section{Introduction}

Exciton-polaritons in solid state microcavities are an active field of research thanks to their potential for both fundamental (Bose Einstein condensates\cite{Kasprzak2006,Plumhof2014,Daskalakis2014}, light superfluidity\cite{Amo2009,Lerario2017}) and practical applications (transistors\cite{Zasedatelev2019}, exciton-polariton lasers\cite{Bhattacharya2013,Kena-Cohen2010}, light-emitting diodes\cite{Mazzeo2014}). Since their first observation\cite{Weisbuch1992} in a planar microcavity, in which an excitonic medium is sandwiched between two mirrors in order to couple the quantized light field, a wide variety of materials including III-V\cite{Weisbuch1992} and II-VI\cite{Kelkar1995} inorganic semiconductors, semiconducting conjugated polymers\cite{Lidzey1998}, small molecules \cite{Holmes2004}, perovskite\cite{Fujita1998} and 2D-materials\cite{Liu2015} have been used for their production.
These strongly-coupled microcavities contain layers of material whose optical refractive indices, which are shaped by the physical properties of the excitons, are either isotropic\cite{Holmes2004}, present an in-plane/out-of-plane uniaxial anisotropy\cite{Kena-Cohen2013}, or an in-plane uniaxial anisotropy for example directed along crystalline axes\cite{Kena-Cohen2008}.

Recently, 1D materials, where the excitons are oriented in a given direction have taken in-plane uniaxial anisotropy inside microcavties to a new extreme as the whole dielectric permittivity becomes concentrated along the director. Demonstrations using oriented nanotubes (made of carbon\cite{Gao2018} or tungsten disulfide\cite{Yadgarov2018}), liquid crystal molecules\cite{Hertzog2017} and liquid crystalline conjugated polymers\cite{LeRoux2019} have been reported. These structures are promising both for enhancing the strength of the coupling inside the microcavity \cite{LeRoux2019,Hertzog2017} and offering a complete contrast between polarizations for novel devices\cite{LeRoux2019}. 

Fitting the energy dispersions of the minima (maxima) of the reflected (transmitted) intensity spectra obtained by varying the polar angle ($\theta$) is usually sufficient to characterize the strength of the interaction inside a microcavity, as increasing $\theta$ is equivalent to increasing the cavity mode energy, which close to the excitonic resonance splits into two extrema, the lower and upper polaritons (LP, UP) separated by the Rabi-splitting energy $\hbar\Omega_{\rm R}$. However, spectral characteristics of cavities containing oriented 1D excitons or in-plane uniaxial anisotropies also depend greatly on the azimuthal angle ($\phi$). 

In this work, we examine the reflectivity and transmissivity of a TE-polarized wave incident on a  microcavity containing  strongly coupled in-plane 1D excitons oriented along the ${\bold e}_{y}$ direction. We focus on the effect of rotating the electric field $\bold{E}$ with respect to the exciton orientation and show that in the case where the cavity damping is broad enough compared with $\hbar\Omega_{y}$ (the Rabi-splitting energy induced by the excitons in the ${\bold e}_{y}$ direction) the measured quantities can closely resemble the ones obtained from a strongly coupled system. In that case, we observe two extrema getting closer in energy as $\phi$ is increased but underline that they are in fact the superposition of separate intensities originating from the propagation of two dephased waves experiencing either the permittivity brought about by the excitons in the ${\bold e}_{y}$ direction (yielding the LP and UP extrema) or the background permittivity in the ${\bold e}_{x}$ direction (yielding a photonic mode extremum). When $\hbar\Omega_{y}$ is much larger than the cavity damping, we recover signals in which the three extrema become clearly resolved and do not experience any spectral shift upon increase of $\phi$. We support our analytical model by fabricating and measuring the TE-reflectivity from a metallic microcavity containing an oriented layer of liquid-crystalline PFO.

\section{TE-polarized Wave Propagation}

\begin{figure}[H]
\includegraphics[scale=0.13]{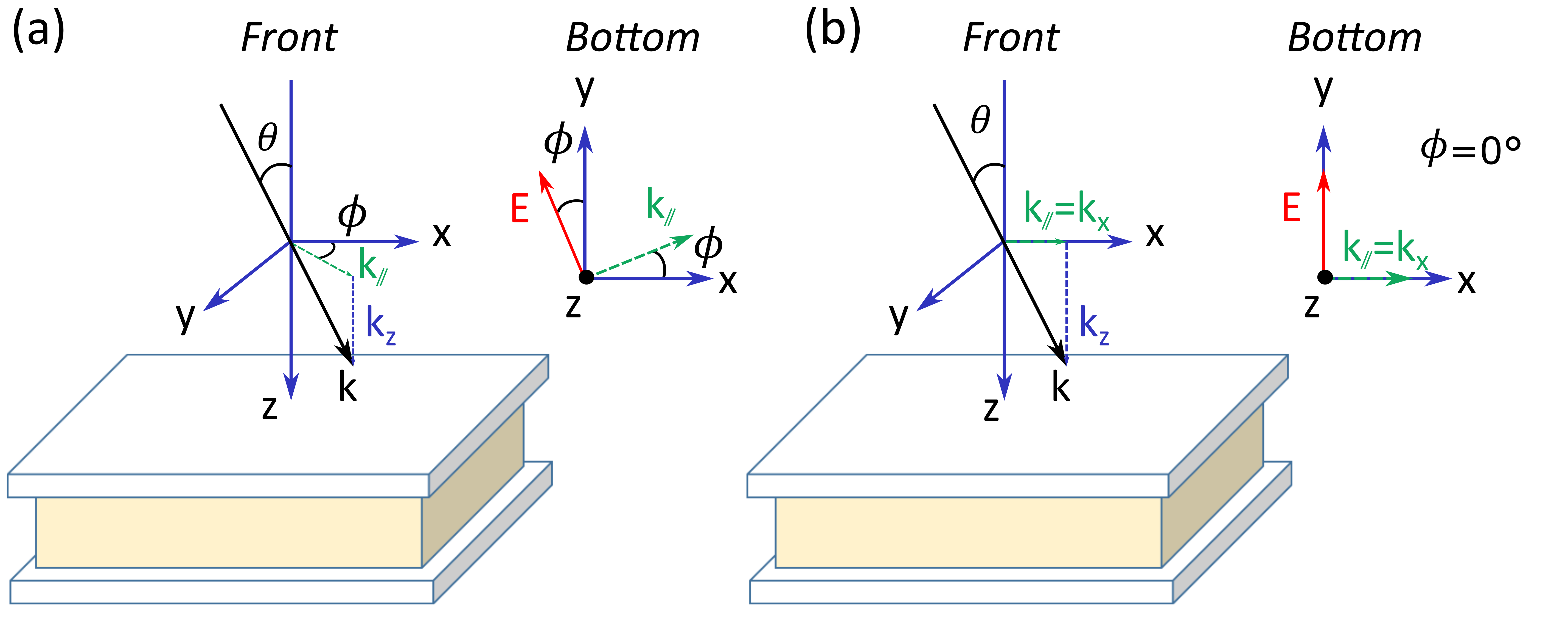}
\caption{\label{fig:FigureGeometry} Geometry and angles used in the main text, $\theta$ is the polar angle formed between $\bold{k}$ and ${\bold e}_{z}$, for a TE-polarized wave $\phi$ is the azimuthal angle formed between $\bold{E}$ and ${\bold e}_{y}$. In (a) $\phi \neq 0^{\circ}$, in (b) $\phi = 0^{\circ}$: $\bold{E}$ is parallel to the transition dipole moment of the excitons and the wave propagates experiencing $\epsilon_{y}$, a similar situtation arises for $\phi = 90^{\circ}$, where $\bold{E}$ becomes parallel to ${\bold e}_{x}$ and the wave experiences $\epsilon_{x}$.}
\end{figure}

We first examine the propagation of an incident TE-polarized wave inside a microcavity containing 1D excitons oriented along ${\bold e}_{y}$. The geometry is shown in Figure~\ref{fig:FigureGeometry}. 

The dielectric permittivity tensor of the cavity layer takes the form:
\begin{equation}\label{eq:DiPermittivity}
\epsilon(\omega)= \epsilon_{0}
\begin{pmatrix}
\epsilon_{x}& 0 & 0\\ 
0 & \epsilon_{y} & 0\\
0 & 0 & \epsilon_{z}\\
\end{pmatrix}   ,
\end{equation}
An incident TE-polarized plane wave on the microcavity can be written:
\begin{equation}\label{eq:propagatingWave}
{\bold E}_{i}(\bold{r},t) = {\bold E}_{0_{i}}{\rm e}^{-{\rm j}[k_{x}x+k_{y}y+k_{z}z -\omega t]} ,
\end{equation}
where:
${\bold E}_{0_{i}} = E_{0_{i}x} {\bold e}_x + E_{0_{i}y} {\bold e}_y$.
The propagation of TE-polarized waves in birefringent media is well documented\cite{Scharf2006,Orfanidis,Yeh1979}: if ${\bold E_{0_{i}}}$ is parallel to either ${\bold e}_{x}$ or ${\bold e}_{y}$ then it propagates experiencing the corresponding axis permittivity (see Figure~\ref{fig:FigureGeometry} (b)), otherwise the $x$ and $y$ components propagate separately and are dephased according to the in-plane anisotropy of the medium (see Supporting Information for more details).
 
We now focus on angle-resolved transmissivity and reflectivity which allow the characterization of the strong coupling regime. We take below the example of reflectivity but note that the reasoning and therefore results are similar for transmissivity. We use a Lorentzian model to represent the electric susceptibility of the excitons inside the cavity layer:
\begin{equation}\label{eq:electric susceptibility}
\chi(\omega) = \frac{4g_{0}^{2}}{\omega_{0}^{2} - \omega^{2} - {\rm j}\gamma\omega},
\end{equation}
where $g_{0}$ is the amplitude of the resonance, $\omega_{0}$ is the pulse frequency of the exciton and $\gamma$ is the natural homogeneous broadening. The dielectric permittivity then reads:
\begin{equation}\label{eq:DiPermittivityUnderModel}
\epsilon(\omega)= \epsilon_{0}
\begin{pmatrix}
\epsilon_{m} & 0 & 0\\ 
0 & \epsilon_{m}(1+\chi(\omega)) & 0\\
0 & 0 & \epsilon_{m}\\
\end{pmatrix}   ,
\end{equation} 
where $\epsilon_{m}$ is the background permittivity of the dielectric layer. We decompose the incoming wave along the ${\bold e}_{x}$ and ${\bold e}_{y}$ directions:
\begin{equation}
\bold{E_{0_{i}}} = \left | \bold{E_{0_{i}}} \right |\begin{pmatrix}
-\sin \phi
\\ 
\cos \phi
\\ 
0
\end{pmatrix}
\end{equation}
On the one hand, the ${\bold e}_{x}$ component experiences a permittivity $\epsilon_{x} = \epsilon_{0}\epsilon_{m}$ and the resulting wave is weakly coupled to the structure, yielding one photonic mode. On the other hand, the ${\bold e}_{y}$ component experiences a permittivity $\epsilon_{y}$ and the physics reverts to that of strong coupling inside an in-plane/out-of-plane medium yielding the LP and UP (see Supporting Information for details). The overall reflected electric field can then be written under the form:
\begin{equation}
\bold{E_{r}} = \left | \bold{E_{0_{i}}} \right |\begin{pmatrix}
-r_{x}(\theta,\omega)\sin \phi
\\ 
r_{y}(\theta,\omega)\cos \phi
\\ 
0
\end{pmatrix}
\end{equation}
where $r_{x}(\theta,\omega)$ and $r_{y}(\theta,\omega)$ can be fully determined using Transfer Matrix (TM) reflectivity calculations. Since $r_{x}(\theta,\omega)$ and $r_{y}(\theta,\omega)$ are calculated using $\epsilon_{x}$ and $\epsilon_{y}$, they do not yield in-phase reflected waves except in very specific cases where the dimensions of the cavity  allow matching of the phase-changes experienced in the two directions. Following averaging, the total reflected intensity becomes the sum of the weighted contributions in the ${\bold e}_{x}$ and ${\bold e}_{y}$ directions:
\begin{equation}
{\bold R_{\bold r}}(\phi,\theta,\omega) = \left | r_x(\theta,\omega) \right |^2\sin^2\phi + \left | r_y(\theta,\omega) \right |^2\cos^2\phi
\end{equation}
which, as we will see in the next section, can lead to the existence of one to three minima (maxima for transmissivity) that require careful interpretation.

\section{Simulation}

We have identified two main cases: either $\hbar\Omega_{y}$ is intense enough so that the LP and UP extrema in reflectivity/transmissivity induced by the ${\bold e}_{y}$ direction are distant enough spectrally to not mix with the central photonic mode extremum induced by the losses of the microcavity in the ${\bold e}_{x}$ direction, which results in three peaks for $\phi \in ]0^{\circ}, 90^{\circ}[$, or the photonic mode broadening along ${\bold e}_{x}$ is comparable to $\hbar\Omega_{y}$ and the three extrema mix to form two peaks for $\phi \in ]0^{\circ}, 90^{\circ}[$ that converge towards the photonic mode energy as $\phi$ is increased.
  
\begin{figure}[H]
\includegraphics[scale=0.15]{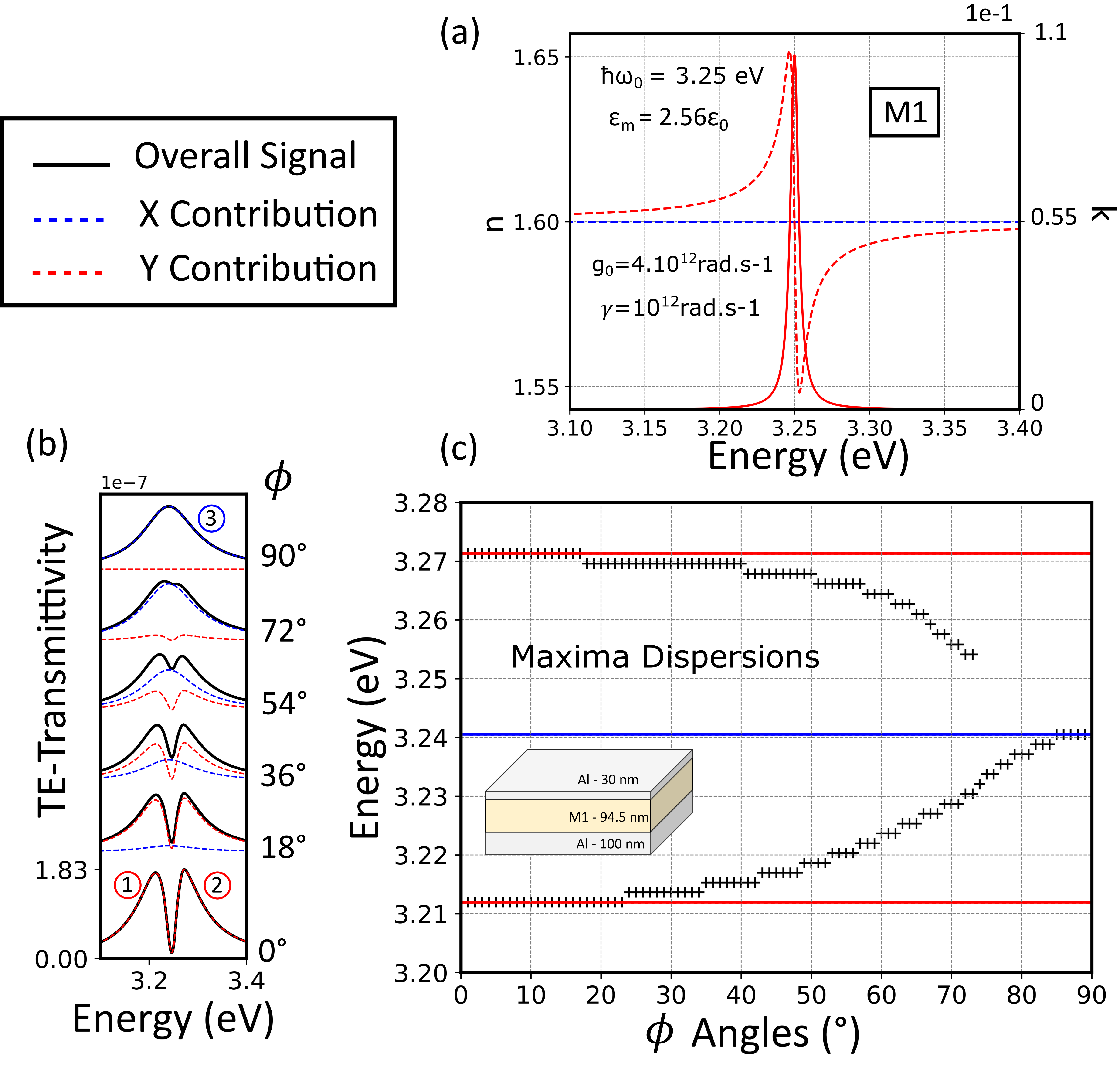}
\caption{\label{fig:FigureSimtransmissivityM1} (a) In red and blue respectively ($n_y$, $k_y$) and ($n_x$, $k_x$) optical components for a simulated material M1 with susceptibility $\chi(\omega)$ using the physical parameters in the inset. Dashed lines give the real component of the complex refractive index $\tilde{n} = n + {\rm i}k$, solid lines the imaginary component. (b) Simulated TE-transmissivity at $\theta = 0^{\circ}$ for $\phi = 0^{\circ}, 18^{\circ}, 36^{\circ}, 54^{\circ}, 72^{\circ}, 90^{\circ}$ of the microcavity displayed in the inset of (c), the black curves represents the total transmissivity which is a superposition of the transmissivities originating from the ${\bold e}_{y}$ direction (in dashed red) and ${\bold e}_{x}$ direction (in dashed blue). Peaks 1 and 2 (in dashed red) are the result of strong light matter coupling in the ${\bold e}_{y}$ direction, peak 3 is the photonic mode (in dashed blue) resulting from the uncoupled ${\bold e}_{x}$ direction. Note the increasing amplitude of peak 3 and decreasing amplitude of peaks 1 and 2 as $\phi$ is increased. (c) Dispersions of the TE-transmissivity maxima for $\phi$ varying by steps of $1^{\circ}$: the black crosses indicate the positions of the maxima for the two peaks in the main signal up until $\phi = 73^{\circ}$, angle from which the two maxima are no longer resolved and replaced by a single maximum, in red the angle-independent positions of the two peaks induced by strong coupling in the ${\bold e}_{y}$ direction, in blue the angle-independent position of the photonic mode in the ${\bold e}_{x}$ direction. Note that the maxima from the overall signal converge towards the photonic mode energy with increasing $\phi$.}
\end{figure}

Figure~\ref{fig:FigureSimtransmissivityM1} (b) shows the simulated transmissivity at normal incidence ($\theta = 0^{\circ}$) for an aluminium microcavity (Al thickness 100 nm at the bottom, 30 nm at the top) containing a 94.5 nm thick layer of material M1 whose susceptibility parameters are: $\epsilon_{m} = 2.56$, $\hbar\omega_{0} = 3.25 \; \rm eV$,  $g_{0} = 4 \times 10^{12} \; \rm rad.s^{-1}$ and $\gamma = 10^{12} \; \rm rad.s^{-1}$ (corresponding to a full width at half maximum (FWHM) $\hbar\gamma \sim 0.7 \; \rm meV$). The cavity mode at normal incidence is slightly detuned $\hbar\Delta = \hbar(\omega_{0} - \omega_{\rm cav}) = 9.5 \; \rm meV$ and its broadening is: $\hbar\kappa = 110 \; \rm meV$. The linewidths of the polaritons in the ${\bold e}_{y}$ direction are measured : $ \rm FWHM_{\rm UP/ \rm LP} \sim 55 \; \rm meV$ which is correctly predicted by: $\rm FWHM_{\rm UP/ \rm LP} = \hbar(\kappa + \gamma)/2$ \cite{Zhu1990,Raizen1989}. Following fitting (see Supporting Information), we derive: $\hbar\Omega_{y} = 58 \; \rm meV$ which is resolved as $\hbar\Omega_{y} > \hbar(\kappa + \gamma)/2$. Since $\hbar\Omega_{y} + \rm FWHM_{\rm UP/\rm LP} < \hbar\kappa$, the transmissivity is composed of two peaks which are the sum of the two polaritons and photonic mode transmissivities mixed together thanks to the large value of $\hbar\kappa$. As $\phi$ is increased, the relative transmissivity of the photonic mode along ${\bold e}_{x}$ increases while the transmissivities of the polaritons along ${\bold e}_{y}$ decrease bringing the peaks closer until they give way to the cavity mode at $3.2405 \; \rm eV$ for $\phi = 90^{\circ}$. The extrema dispersions are represented in Figure~\ref{fig:FigureSimtransmissivityM1} (c). We note that each transmissivity maximum is the result of contributions from different out-of phase waves in the ${\bold e}_{x}$ and ${\bold e}_{y}$ directions for $\phi \in ]0^{\circ}, 90^{\circ}[$ and as such does not characterize an eigenmode of the structure.

\begin{figure}[H]
\includegraphics[scale=0.15]{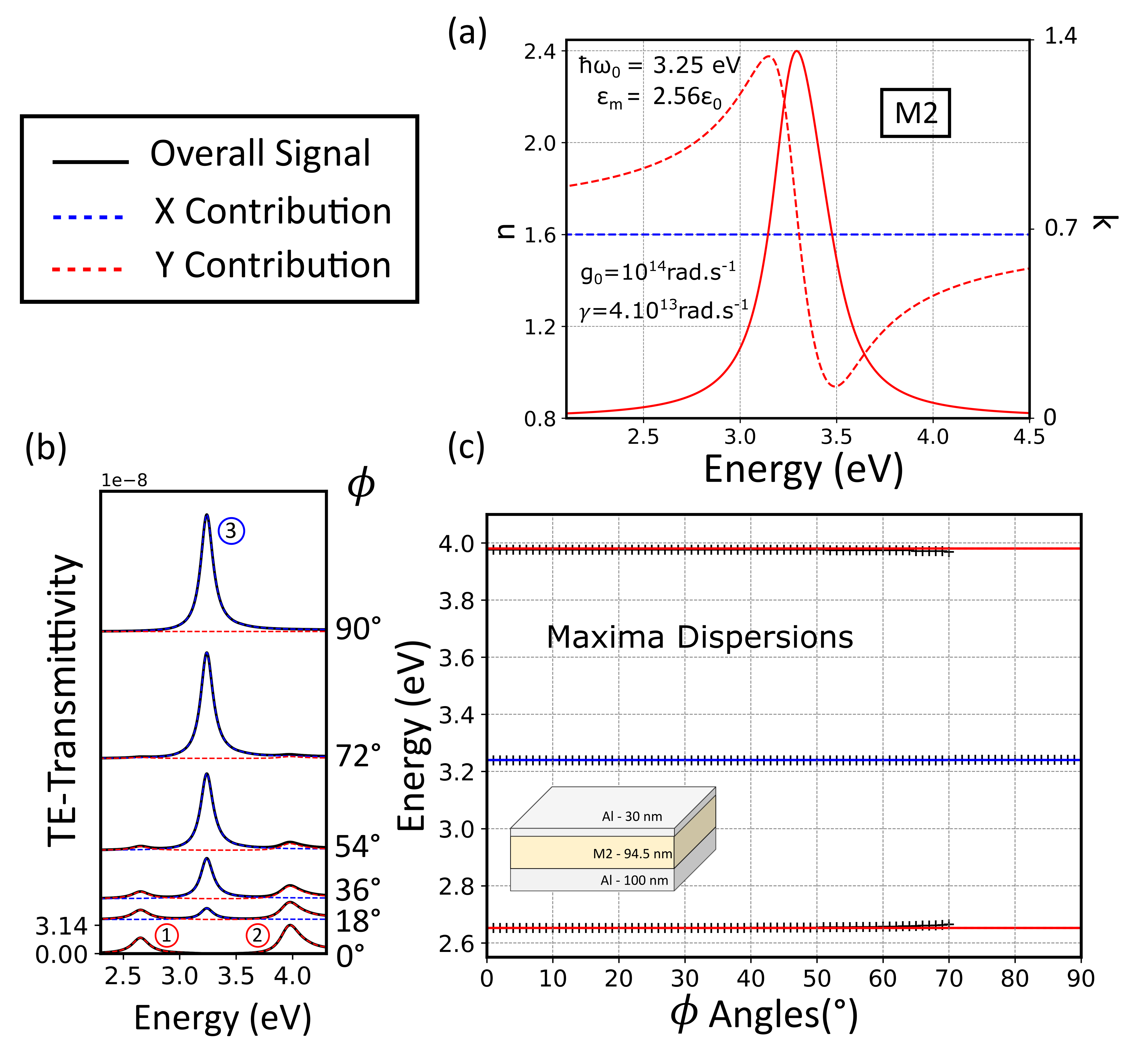}
\caption{\label{fig:FigureSimtransmissivityM2} (a) In red and blue respectively ($n_y$, $k_y$) and ($n_x$, $k_x$) optical components for a simulated material M2 with susceptibility $\chi(\omega)$ using the physical parameters in the inset. Dashed lines give the real component of the complex refractive index $\tilde{n} = n + {\rm i}k$, solid lines the imaginary component. (b) Simulated TE-transmissivity at $\theta = 0^{\circ}$ for $\phi = 0^{\circ}, 18^{\circ}, 36^{\circ}, 54^{\circ}, 72^{\circ}, 90^{\circ}$ of the microcavity displayed in the inset of (c), the black curves represents the total transmissivity which is a superposition of the transmissivities originating from the ${\bold e}_{y}$ direction (in dashed red) and ${\bold e}_{x}$ direction (in dashed blue). Peaks 1 and 2 (in dashed red) are the result of strong light matter coupling in the ${\bold e}_{y}$ direction, peak 3 is the photonic mode (in dashed blue) resulting from the uncoupled ${\bold e}_{x}$ direction. Note the increasing amplitude of peak 3 and decreasing amplitude of peaks 1 and 2 as $\phi$ is increased. (c) Dispersions of the TE-transmissivity maxima for $\phi$ varying by steps of $1^{\circ}$: the black crosses indicate the angle-independent positions of the maxima for the well resolved three peaks in the main signal up until $\phi = 70^{\circ}$, angle from which the two maxima from the ${\bold e}_{y}$ direction are no longer visible, in red the angle-independent positions of the two peaks induced by strong coupling in the ${\bold e}_{y}$ direction, in blue the angle-independent position of the photonic mode in the ${\bold e}_{x}$ direction. Note that the positions of the maxima in the main signal matches the positions of peak 1, 2 and 3.}
\end{figure}

Figure~\ref{fig:FigureSimtransmissivityM2} (b) shows the simulated transmissivity at normal incidence ($\theta = 0^{\circ}$) for an aluminium microcavity (Al thickness 100 nm at the bottom, 30 nm at the top) containing a 94.5 nm thick layer of material M2 whose susceptibility parameters are: $\epsilon_{m} = 2.56$, $\hbar\omega_{0} = 3.25\,$eV,  $g_{0} = 10^{14} \,$rad.s$^{-1}$ and $\gamma = 4 \times 10^{13} \,$ rad.s$^{-1}$ (corresponding to a full width at half maximum $\rm FWHM = \hbar\gamma \sim 26 \,\rm meV$). For this structure, $\hbar\Omega_{y}$ is fitted: $\hbar\Omega_{y} = 1.38 \, \rm eV$. This splitting represents $\sim 42 \%$ of the exciton energy $\hbar\omega_{0} = 3.25 \, \rm eV$ bringing the system into the ultrastrong coupling regime (USC) in the ${\bold e}_{y}$ direction. $\hbar\kappa$ is this time more than one order of magnitude smaller than $\hbar\Omega_{y}$ and $\hbar\Omega_{y}  + \rm FWHM_{\rm UP/\rm LP} \gg \hbar\kappa$, the transmissivity now resolves three peaks for $\phi \in ]0^{\circ},90^{\circ}[$ which are the two polaritons and the photonic mode transmissivities weighted by $\phi$: as $\phi$ is increased, the relative transmissivity of the photonic mode along ${\bold e}_{x}$ increases while the transmissivities of the polaritons along ${\bold e}_{y}$ decrease changing the relative heights of the peaks without spectrally shifting them. The extrema positions are represented in Figure~\ref{fig:FigureSimtransmissivityM1} (c).

\begin{figure}[H]
\includegraphics[scale=0.26]{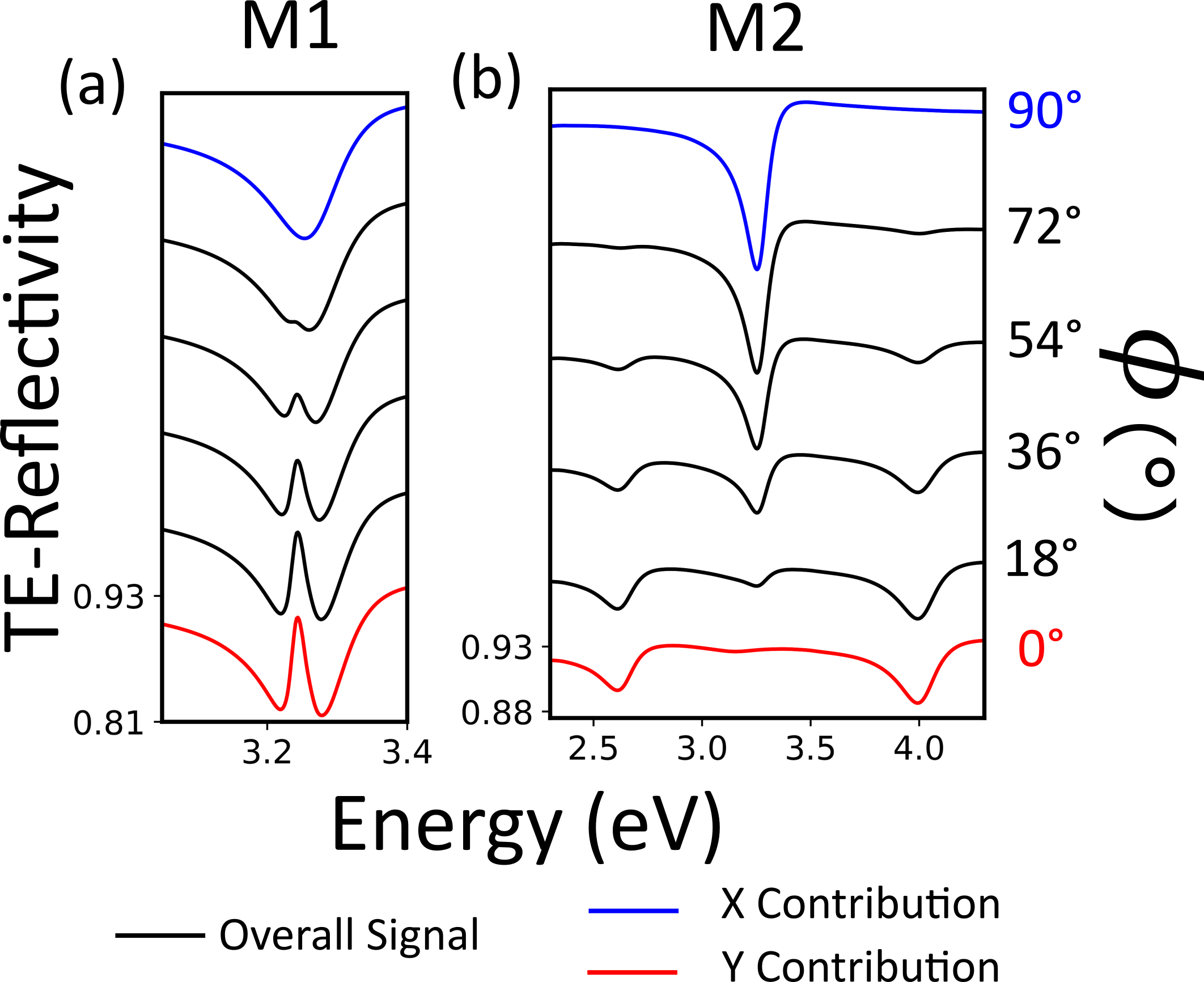}
\caption{\label{fig:FigureSimReflectivity} Simulated TE-reflectivity at $\theta = 30^{\circ}$ for $\phi = 0^{\circ}, 18^{\circ}, 36^{\circ}, 54^{\circ}, 72^{\circ}, 90^{\circ}$ of the microcavities displayed in the inset of Figure~\ref{fig:FigureSimtransmissivityM1}(c) for (a) and Figure~\ref{fig:FigureSimtransmissivityM2}(c) for (b) with a 99.5 nm material layer thickness. In both cases the simulated values at $\phi = 0^{\circ}$ (in solid red) correspond to the reflectivity contributions from the LP and UP in the ${\bold e}_{y}$ direction, the simulated values at $\phi = 90^{\circ}$ (in solid blue) correspond to the reflectivity contribution from the photonic mode in the ${\bold e}_{x}$ direction. For other values of $\phi$, two main minima are visible in (a) and get closer in energy with increasing values of $\phi$ as the relative reflectivity contribution along ${\bold e}_{y}$ decreases and the one along ${\bold e}_{x}$ increases until only one peak corresponding to the photonic mode ${\bold e}_{x}$ is visible. In (b) three main minima are visible but do not shift in energy with increasing values of $\phi$ as the relative reflectivity contribution along ${\bold e}_{y}$ decreases and the one along ${\bold e}_{x}$ increases.}
\end{figure}

We note that the TE-transmissivities obtained in Figure~\ref{fig:FigureSimtransmissivityM1} (b) and Figure~\ref{fig:FigureSimtransmissivityM2} (b) are too low to be actually measured. Figure~\ref{fig:FigureSimReflectivity} (a) and (b) represent simulated TE-reflectivities for similar structures differing in a 99.5 nm thick cavity layer and the measurement being performed at $\theta = 30^{\circ}$ to keep the detuning at $\Delta = 10 \; \rm meV$. The origin of the minima observed are analog to the maxima in transmissivity and the interpretation is identical to the one previously made. We then proceed in the following section to fabricate and measure a microcavity with similar characteristics to the one showed in Figure~\ref{fig:FigureSimtransmissivityM2} (c).

\section{Experiment}

\begin{figure}[H]
\includegraphics[scale=0.20]{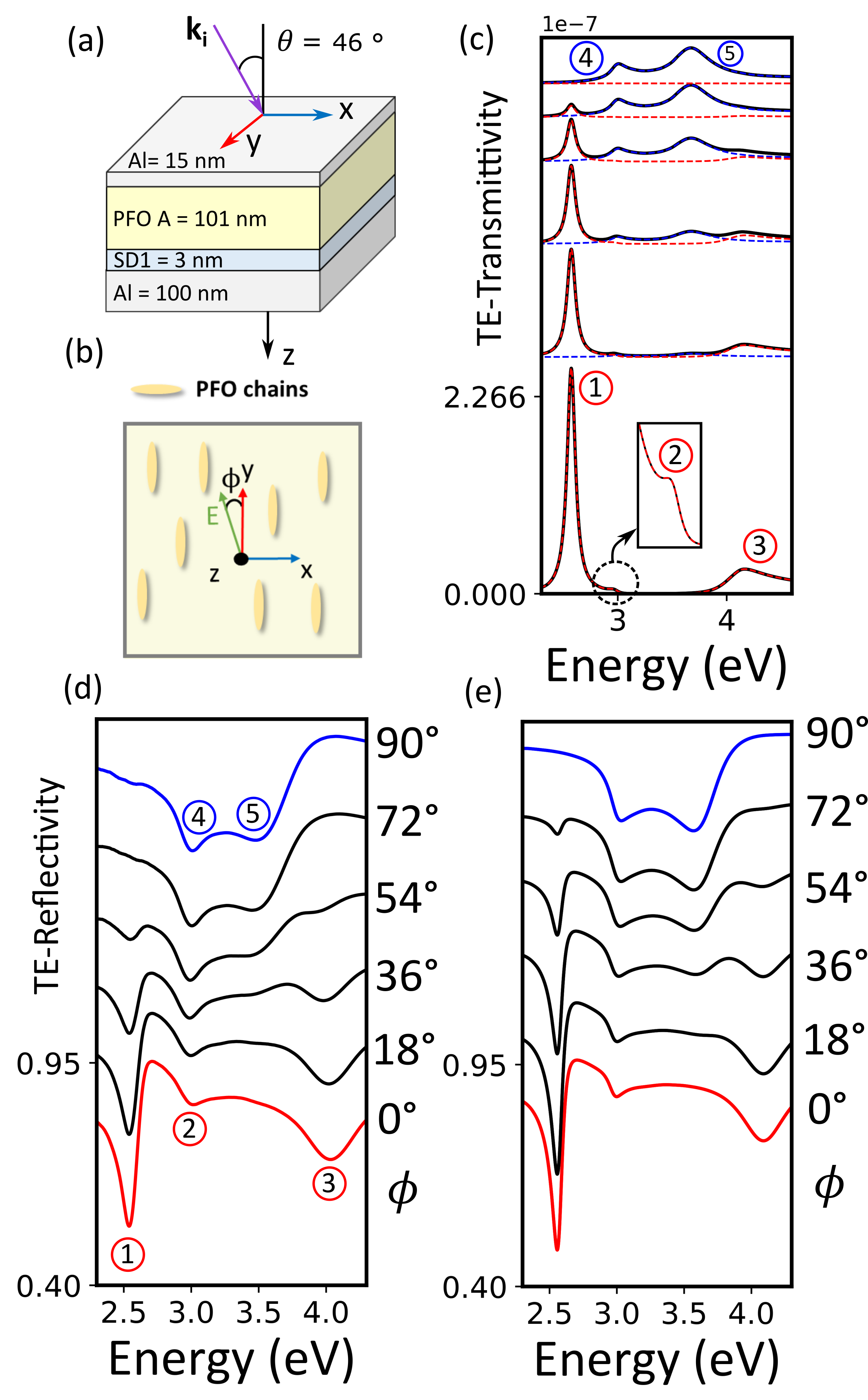}
\caption{\label{fig:FigureExperiment} (a) Schematic of the microcavity structure fabricated. The PFO chains were oriented (A) using a photoaligned SD1 layer: the general fabrication process is described in Ref 21. The TE-reflectivity is measured at $\theta = 46^{\circ}$. (b) Bottom view of the orientation of the PFO chains along the ${\bold e}_{y}$ direction, the measurement is performed by rotating the sample relative to the electric field $\bold{\rm E}$ (in green), forming the azimuthal angle $\phi$. (c) Simulated TE-transmissivity at $\theta = 46^{\circ}$ for $\phi = 0^{\circ}, 18^{\circ}, 36^{\circ}, 54^{\circ}, 72^{\circ}, 90^{\circ}$, the black curves represent the total transmissivity which is a superposition of the transmissivities originating from the ${\bold e}_{y}$ direction (in dashed red) and ${\bold e}_{x}$ direction (in dashed blue). Peaks 1, 2 and 3 (in dashed red) are the result of ultra strong light matter coupling in the ${\bold e}_{y}$ direction for the lowest lying energy cavity mode (1 and 3) and the second lowest lying energy cavity mode (2). Peak 4 and 5 are the result of a splitting of the photonic mode in the ${\bold e}_{x}$ direction due to remaining optical activity. Further explanations can be found in text. (d) Measured TE-reflectivity at $\theta = 46^{\circ}$ for $\phi = 0^{\circ}, 18^{\circ}, 36^{\circ}, 54^{\circ}, 72^{\circ}, 90^{\circ}$, the numbered minima correspond to the ones simulated in transmissivity. (e) Simulated TE-Reflectivity at $\theta = 46^{\circ}$ for $\phi = 0^{\circ}, 18^{\circ}, 36^{\circ}, 54^{\circ}, 72^{\circ}, 90^{\circ}$ confirming the experimental observations.}
\end{figure}

We fabricated a microcavity containing a layer of PFO whose chains were oriented along ${\bold e}_{y}$ thanks to the use of a photoalignment-layer-induced homogeneous nematic orientation technique \cite{Zhang2019,LeRoux2019}, the structure and orientation are shown in Figure~\ref{fig:FigureExperiment} (a) and (b), the exact fabrication process is detailed elsewhere \cite{LeRoux2019}. The refractive index of the oriented PFO is reproduced from Ref 21 and shown in Supporting Information, the TE-reflectivity was measured for different values of $\phi$ and $\theta$ thanks to the experimental setup described in Ref 21. The coupling strength for $\bold{E}$ parallel to ${\bold e}_{y}$ was previously determined: $\hbar\Omega_{y} = 1.47 \; \rm eV$\cite{LeRoux2019} and is found to be identical here (see Supporting Information for the fittings). TM reflectivity calculations confirm the dimensions of the structure and from both simulations and experiments, zero detuning between cavity mode and exciton (centered at $3.25 \; \rm eV$) is reached for $\theta = 46^{\circ}$. 

Compared with the simulations in the previous section, the large broadening of the exciton distribution is inhomogeneous and is the result of deviations from the mean fluorene-fluorene single bond torsion angle ($\sim 135^{\circ}$) that generates multiple conformers causing inhomogeneous broadening of the main $S_{0}-S_{1}$ absorption\cite{Chunwaschirasiri2005}. Inhomogeneous broadening was shown not to alter $\hbar\Omega$\cite{Houdre1996,Savona1996,Wang1997} as long as its value is modest compared with the splitting itself and as such does not have a fundamental impact on our analysis.

Figure~\ref{fig:FigureExperiment} (c) shows the simulated TE-transmissivity for $\theta = 46^{\circ}$ for increasing values of $\phi = 0^{\circ}, 18^{\circ}, 36^{\circ}, 54^{\circ}, 72^{\circ}, 90^{\circ}$. At $\phi = 0^{\circ}$, $\bold{E}$ is parallel to ${\bold e}_{y}$ and we observe two main maxima (1 and 3) that correspond to the LP and UP. Interestingly, we observe the formation of another peak (2) close to the LP at $\sim 2.95 \; \rm eV$ which corresponds to a further lower polariton brought about by the strong coupling of the second lowest lying cavity mode with the excitons (this coupling is made possible thanks to the intense absorption in the ${\bold e}_{y}$ direction), this shoulder peak is shown in more detail in the inset of Figure~\ref{fig:FigureExperiment} (c).

As $\phi$ is increased, we observe the formation of four maxima. The two most outer ones in energy correspond to the LP and UP transmittivities 1 and 3 created by the strong coupling of the lowest lying cavity mode to the excitons: given the large value of $\hbar\Omega_{y} = 1.47 \; \rm eV$ compared with the cavity mode broadening $\hbar\kappa = 220  \; \rm meV$ those are similar in nature to the outer maxima in Figure~\ref{fig:FigureSimtransmissivityM2} (b) and consequently do not shift in energy as $\phi$ is increased. The two remaining maxima are better understood by examining $\phi = 90^{\circ}$: in that case, $\bold{E}$ is perpendicular to ${\bold e}_{y}$ and the remaining optical activity along ${\bold e}_{x}$ (which is not simply the background permittivity $\epsilon_{m}$ compared with Equation~\ref{eq:DiPermittivityUnderModel}) causes the photonic mode to split around the exciton at 3.25 eV (peak 4 at 3.0 eV and peak 5 at 3.67 eV). Maxima 4 and 5 are however not clearly resolved since the oscillator strength along ${\bold e}_{x}$ is not intense enough following orientations of the PFO chains: it is an illustration of the switchable coupling strength that exists between the two orthogonal directions\cite{Hertzog2017,LeRoux2019}. As for the main signal, for $\phi \in ]0^{\circ}, 90^{\circ}[$ the lowest energy central maximum is the superposition of peak 2 and 4 which lie 0.05 eV apart in energy and whose respective broadenings allow for their mixing, making the overall maximum seemingly shift to lower energies as $\phi$ is increased. Finally, the last main maximum corresponds to peak 5 and does not shift with increasing $\phi$.
 
Figure~\ref{fig:FigureExperiment} (d) shows the experimental measurement of the TE-reflectivity at $\theta = 46^{\circ}$. All the minima are similar to the maxima identified using the simulated TE-transmissivity and their spectral positions is confirmed by simulating the TE-reflectivity in Figure~\ref{fig:FigureExperiment} (e).

\section{Conclusion}
We have carefully examined the effects of rotating the electric field of an incident TE-polarized wave on the measured reflectivity and transmissivity for a microcavity containing in-plane oriented, strongly coupled 1D excitons. We have demonstrated that when the cavity damping $\hbar\kappa$ is broad enough compared to $\hbar\Omega_{y}$, TE-transmissivity and reflectivity present two extrema for $\phi \in ]0^{\circ}, 90^{\circ}[$ which get closer in energy as $\phi$ is increased. While this resembles the result obtained for a strongly coupled microcavity upon increase of the polar angle $\theta$, these extrema are the superposition of separate intensities originating from the propagation of two dephased waves experiencing either strong coupling in the ${\bold e}_{y}$ direction or weak coupling in the ${\bold e}_{x}$ direction but mixed thanks to the losses $\hbar\kappa$. In the case where $\hbar\Omega_{y}$ is much larger than the cavity damping, we showed that the measured reflectivity/transmissivity present three extrema for $\phi \in ]0^{\circ}, 90^{\circ}[$ which are the separate contributions from the different directions, as such they do not experience any spectral shifting with increasing $\phi$. We supported our analytical model by fabricating and measuring the TE-reflectivity from a metallic microcavity containing an oriented layer of liquid-crystalline PFO and interpreted the different extrema observed by considering separately the contributions from the ${\bold e}_{x}$ and ${\bold e}_{y}$ directions. We believe that this work will help the development of applications based on microcavities containing in-plane oriented strongly coupled 1D excitons.

\begin{acknowledgement}

The authors acknowledge funding from the University of Oxford, from the UK Engineering and Physical Sciences Research Council and the Jiangsu Industrial Technology Research Institute. F.L.R. further thanks Wolfson College and Dr Simon Harrison for the award of a Wolfson Harrison UK Research Council Physics Scholarship.

\end{acknowledgement}

\bibliography{manuscriptbiblio}

\end{document}

\newpage

\begin{figure}[H]
\renewcommand{\thefigure}{S1}
\includegraphics[scale=0.13]{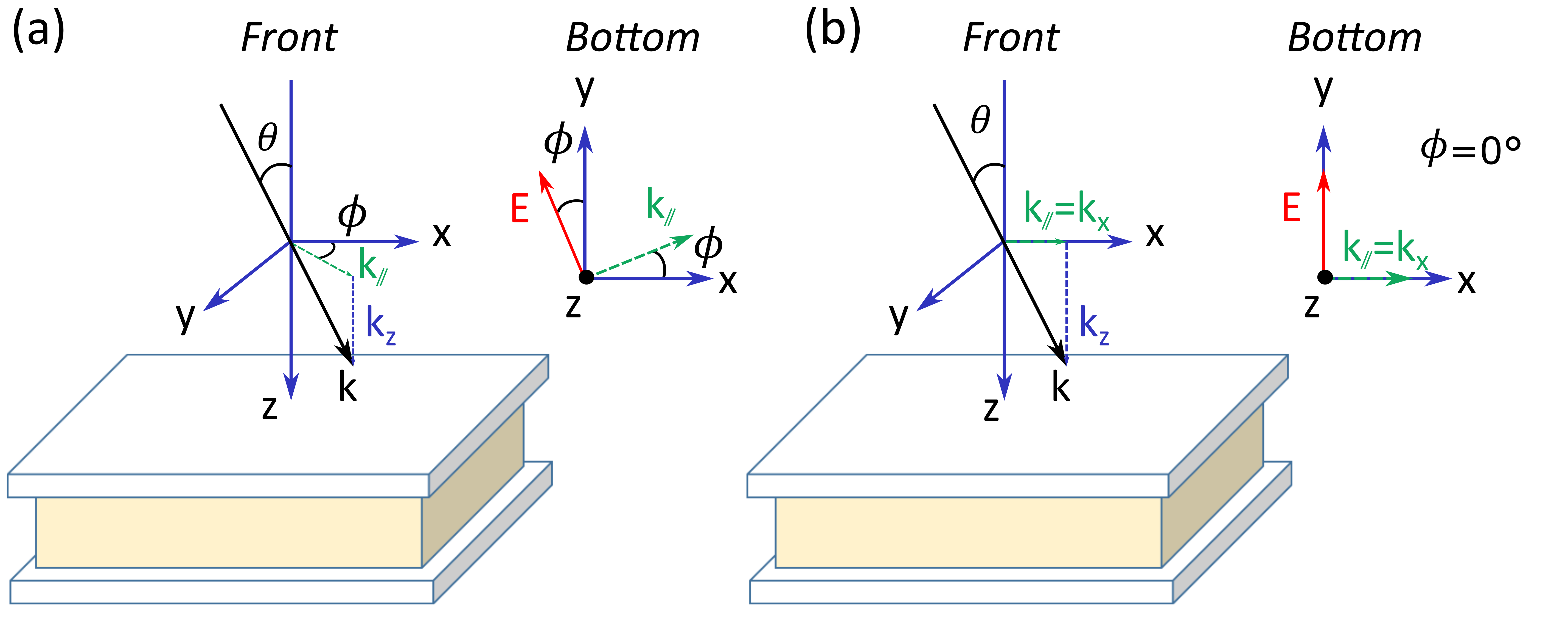}
\caption{\label{fig:FigureGeometry} Geometry and angles used in the main text, $\theta$ is the polar angle formed between $\bold{k}$ and ${\bold e}_{z}$, for a TE-polarized wave $\phi$ is the azimuthal angle formed between $\bold{E}$ and ${\bold e}_{y}$. (a) represents the general case where $\phi \neq 0^{\circ}$, (b) represents the case where $\phi = 0^{\circ}$ which can always be used when the structure presents a rotational symmetry around ${\bold e}_{z}$.}
\end{figure}

The different angles and general geometry of the problem are shown in Figure~\ref{fig:FigureGeometry}.

\section{TE-polarized Wave Propagation in an In-plane/Out-of-plane Anisotropic Medium}

In this section we examine the validity of Snell law and the norm of the wavevector $\bold{k}$ for a TE-polarized wave propagating in a dielectric medium whose dielectric permittivity tensor is diagonal but not proportional to the identity tensor. We focus on the case of an in-plane/out-of-plane anisotropy i.e. $\epsilon_{x} = \epsilon_{y} = \epsilon_{\rm ord}$ and $\epsilon_{z} = \epsilon_{\rm ex}$ where $\epsilon_{\rm ord}$ and $\epsilon_{\rm ex}$ are respectively the ordinary (in-plane) and extraordinary (out-of-plane) components. We recall the Helmholtz wave equation for the different media:
\begin{equation}\label{eq:waveEquation}
\bold{k}\times[(\mu^{-1})\cdot\bold{k}\times\bold{E}] + \frac{\omega^{2}}{c^{2}}\mathit{\epsilon}\cdot\bold{E} = \bold{0},
\end{equation}
where the permeability $\mu$ is assumed to be the identity matrix for all layers. Equation~\ref{eq:waveEquation} can be reformulated using the matrix $\mathit{M(\omega)}$ such that:
\begin{equation}\label{eq:propagatingWaveMatrixEquation}
\mathit{\bold{M}(\omega)}\bold{E} = \bold{0},
\end{equation}
A propagating wave solution of Equation~\ref{eq:waveEquation} incident on the microcavity structure takes the form:
\begin{equation}\label{eq:propagatingWave}
{\bold E}_{i}(\bold{r},t) = {\bold E}_{0_{i}}{\rm e}^{-{\rm j}[k_{x}x+k_{y}y+k_{z}z -\omega t]} ,
\end{equation}
In Equation~\ref{eq:propagatingWaveMatrixEquation} $\bold{M}$ takes the form:
\begin{equation}\label{eq:propagatingWaveMatrixG}
\mathit{\bold{M}(\omega)} = \begin{pmatrix}
\epsilon_{x}\frac{\omega^{2}}{c^{2}} - k_{y}^{2} - k_{z}^{2} & k_{x}k_{y} & k_{x}k_{z} \\ 
k_{y}k_{x} &  \epsilon_{y}\frac{\omega^{2}}{c^{2}} - k_{z}^{2} - k_{x}^{2} & k_{y}k{z}\\ 
k_{z}k_{x} & k_{z}k_{y} & \epsilon_{z}\frac{\omega^{2}}{c^{2}} - k_{x}^{2} - k_{y}^{2}
\end{pmatrix}
\end{equation}
We recall Maxwell-Faraday law for the electric and magnetic fields:
\begin{equation}\label{eq:MaxwellFaraday}
\bold{k}\times\bold{E} = \frac{\omega}{c}\bold{B}
\end{equation}
Finally, we focus on the case of TE-polarized light: ${\bold E}_{0_{i}} = E_{0_{i}x} {\bold e}_x + E_{0_{i}y} {\bold e}_y$.

\subsection{In-plane/Out-of-plane anisotropy}

The dielectric permittivity takes the form:
\begin{equation}\label{eq:UniaxialModelG}
\epsilon(\omega)= \epsilon_{0}
\begin{pmatrix}
\epsilon_{\rm ord}& 0 & 0\\ 
0 & \epsilon_{\rm ord} & 0\\
0 & 0 & \epsilon_{\rm ex}\\
\end{pmatrix}   ,
\end{equation}  
The rotational symmetry around ${\bold e}_{z}$ allows us to choose:
\begin{equation}\label{eq:WaveVectorUniaxialSymmetry}
\bold{k} = \begin{pmatrix}
k_{x}
\\ 0
\\k_{z}
\end{pmatrix}, \, k_{y} = 0
\end{equation}
and
\begin{equation}\label{eq:EFieldUniaxial}
\bold{E}_{0_{i}} = \begin{pmatrix}
0
\\ E_{0_{iy}}
\\ 0
\end{pmatrix}\,{\rm with} \,E_{0_{iy}}\neq 0
\end{equation}
which matches the schematic in Figure~\ref{fig:FigureGeometry} (b) without loss of generality.
Equation~\ref{eq:propagatingWaveMatrixEquation} then reduces to:
\begin{equation}\label{eq:reducedWaveEquationG}
\epsilon_{\rm ord}\frac{\omega^{2}}{c^{2}} - k_{z}^{2} - k_{x}^{2} = 0.
\end{equation}
As $\bold{k^{2}} = k_{x}^{2} + k_{z}^{2}$, we have:
\begin{equation}\label{eq:normUniaxial}
\bold{k^{2}} = \epsilon_{\rm ord}\frac{\omega^{2}}{c^{2}}.
\end{equation}
Using the continuity of the normal component of $\bold{B} : B_{z}$ and the tangential components of $\bold{E} : E_{x}, E{y}$  at the dielectric interfaces and Maxwell-Faraday law (Equation~\ref{eq:waveEquation}) we retrieve:
\begin{equation}\label{eq:SnellUniaxialDerivation}
B_{z_{n-1}} = B_{z_{n}}\Leftrightarrow  
k_{x_{n-1}}E_{y_{n-1}} = k_{x_{n}}E_{y_{n}}
\Leftrightarrow 
k_{x_{n-1}} = k_{x_{n}} 
\end{equation}
We propagate this equation from the vacuum in which the incident wave propagates with an angle $\theta$ to the uniaxial layer in which it propagates with an angle $\theta_{m}$ and retrieve Snell's law:
\begin{equation}\label{eq:SnellUniaxial}
k_{x_{0}}  = k_{x_{n}} \Rightarrow \left | k \right | \sin\theta  = \left | k_{n} \right | \sin\theta_m \Leftrightarrow \sin\theta  = n_{\rm ord} \sin\theta_m
\end{equation}

\subsection{Biaxial Anisotropy}

We study the case where the permittivity tensor has different diagonal elements for all three coordinates $(x,y,z)$: 
\begin{equation}\label{eq:BiTensor}
\epsilon(\omega)= \epsilon_{0}
\begin{pmatrix}
\epsilon_{x}& 0 & 0\\ 
0 & \epsilon_{y} & 0\\
0 & 0 & \epsilon_{z}\\
\end{pmatrix}   ,
\end{equation}
In particular we justify the claim made in the main text that a TE-polarized plane wave can not propagate as a single wave through the cavity medium except if the incoming $\bold{E}_{i}$ is parallel to one of the principal axes ${\bold e}_{x}$ or ${\bold e}_{y}$. Equation~\ref{eq:propagatingWaveMatrixEquation} yields:
\begin{equation}\label{eq:fromLastLineMatrixForm}
\left\{\begin{matrix}
(\epsilon_{x}\frac{\omega^{2}}{c^{2}} - k_{y}^{2} - k_{z}^{2})E_{x} + k_{x}k_{y}E_{y} = 0
\\
(\epsilon_{y}\frac{\omega^{2}}{c^{2}} - k_{z}^{2} - k_{x}^{2})E_{y} + k_{y}k_{x}E_{x} = 0
\\
k_{z}k_{x}E_{x} + k_{z}k_{y}E_{y} = 0
\end{matrix}\right.
\end{equation}
We first examine the case where $\bold{E}_{0_{i}}$ is not parallel to ${\bold e}_{x}$ or ${\bold e}_{y}$ and suppose that the incident wave can propagate through the structure, we have:
\begin{equation}\label{eq:generalCdtBiaxial}
\left\{\begin{matrix}
E_{x} \neq 0 \, \rm and \, \mathit{E_{y}} \neq 0
\\ 
k_{x} \neq 0 \, \rm and \, \mathit{k_{y}} \neq 0
\\
k_{z} \neq 0
\end{matrix}\right.
\end{equation}
As $k^{2} = k_{x}^{2} + k_{y}^{2} + k_{z}^{2}$, we use the third equation in System~\ref{eq:fromLastLineMatrixForm} and obtain: $k_{x}E_{x} = -k_{y}E_{y}$ which we re-introduce in System~\ref{eq:generalCdtBiaxial} to derive the equivalent system:
\begin{equation}\label{eq:fromLastLineMatrixFormSimplified}
\left\{\begin{matrix}
\epsilon_{x}\frac{\omega^{2}}{c^{2}} - k^{2} = 0
\\
\epsilon_{y}\frac{\omega^{2}}{c^{2}} - k^{2} = 0
\\
k_{z}k_{x}E_{x} + k_{z}k_{y}E_{y} = 0
\end{matrix}\right.
\end{equation}
Substracting the second equation to the first equation in System~\ref{eq:fromLastLineMatrixFormSimplified} leads to a contradiction:
\begin{equation}\label{eq:ContradictionBiaxial}
\epsilon_{x}-\epsilon_{y} = 0.
\end{equation}
Therefore an incident TE-polarized wave can not propagate in the dielectric as a single wave if $\bold{E}_{i}$ is not parallel to either ${\bold e}_{y}$ or ${\bold e}_{x}$.

\section{Strong Exciton-Photon Coupling in a Microcavity Containing an In-plane/Out-of-plane Anisotropic Central Layer}

In this section, we recall a semi-classical demonstration of strong exciton photon coupling for a microcavity containing an in-plane/out-of-plane anisotropic layer sandwiched between two mirrors, physically possible since any incident TE-polarized wave can propagate through the structure. The excitons inside the semiconductor layer are modelled using a Lorentzian model for the electric susceptibility:
\begin{equation}\label{eq:electric susceptibility}
\chi(\omega) = \frac{4g_{0}^{2}}{\omega_{0}^{2} - \omega^{2} - {\rm j}\gamma\omega},
\end{equation}
where $g_{0}$ is the amplitude of the resonance, $\omega_{0}$ is the pulse frequency of the exciton resonance and $\gamma$ is the natural homoegeneous broadening. To find the real pulse solutions of the model that follows, we set $\gamma = 0 \; \rm rad.s^{-1}$, reports including non-zero $\gamma$ values are available in the litterature\cite{Zhu1990,Raizen1989} and allow to characterize the linewidths of the derived polaritons. The permittivity tensor then reads:
\begin{equation}\label{eq:UniaxialTensor}
\epsilon(\omega)= \epsilon_{0}
\begin{pmatrix}
\epsilon_{m}(1+\chi(\omega)) & 0 & 0\\ 
0 & \epsilon_{m}(1+\chi(\omega)) & 0\\
0 & 0 & \epsilon_{m}\\
\end{pmatrix}   ,
\end{equation} 
where $\epsilon_{m}$ is the background permittivity of the dielectric layer. From Equation~\ref{eq:propagatingWaveMatrixEquation}, $\bold{M}$ now takes the form:
\begin{equation}\label{eq:propagatingWaveMatrixUni}
\mathit{\bold{M}(\omega)} = \begin{pmatrix}
[1+\chi(\omega)]\epsilon_{m}\frac{\omega^{2}}{c^{2}} - k_{y}^{2} - k_{z}^{2} & k_{x}k_{y} & k_{x}k_{z} \\ 
k_{y}k_{x} &  [1+\chi(\omega)]\epsilon_{m}\frac{\omega^{2}}{c^{2}} - k_{z}^{2} - k_{x}^{2} & k_{y}k{z}\\ 
k_{z}k_{x} & k_{z}k_{y} & \epsilon_{m}\frac{\omega^{2}}{c^{2}} - k_{x}^{2} - k_{y}^{2}
\end{pmatrix}
\end{equation}
We focus on the case of TE-polarized light: ${\bold E}_{0_{i}} = E_{0_{i}x} {\bold e}_x + E_{0_{i}y} {\bold e}_y$ (the TM-polarized light case can be found in the literature\cite{Kena-Cohen2013}). As for the first section, thanks to the rotational symmetry around ${\bold e}_{z}$, $\bold{k}$ takes the same form as in Equation~\ref{eq:WaveVectorUniaxialSymmetry} and $\bold{E}_{i}$ the one in Equation~\ref{eq:EFieldUniaxial}. Equation~\ref{eq:propagatingWaveMatrixEquation} then reduces to:
\begin{equation}\label{eq:reducedWaveEquation}
[1+\chi(\omega)]\epsilon_{m}\frac{\omega^{2}}{c^{2}} - k_{z}^{2} - k_{x}^{2} = 0
\end{equation}
Inside the cavity layer. $k_{z}$ is quantized, using the lowest cavity mode at normal incidence ($\theta = 0^{\circ}$) we have:
\begin{equation}\label{eq:kzQuantization}
k_{z}^{2} = \epsilon_{m}\frac{\omega_{{\rm cav}_{0}}^{2}}{c^{2}}
\end{equation}
Using Equation~\ref{eq:normUniaxial}:
\begin{equation}\label{eq:normKUniaxial}
k^{2} = \epsilon_{m}\frac{\omega_{\rm cav}^{2}}{c^{2}}
\end{equation}
and as Snell law is valid ($\theta_{m}$ being the propagation angle inside the central layer):
\begin{equation}\label{eq:normKUniaxialFullExp}
k^{2}\cos^2\theta_{m} = k_{z}^{2}\Leftrightarrow \omega_{cav}^{2} = \frac{\omega_{{\rm cav}_{0}}^{2}}{1-\frac{\sin^{2}\theta}{\epsilon_{m}}} 
\end{equation}
We re-write Equation~\ref{eq:reducedWaveEquation} using $k_{x}^{2} = k^{2}\sin^{2}\theta_m = \frac{\omega^{2}}{c^{2}}\sin^{2}\theta$ and replace $\chi(\omega)$ by its expression, we obtain:
\begin{equation}\label{eq:polEquations}
4g_{0}^{2} = (\omega^{2} - \omega_{0}^{2})(\omega^{2} - \omega_{\rm cav}^{2})
\end{equation}
When $\omega_{\rm cav} = \omega_{0}$ we solve Equation~\ref{eq:polEquations} to find: $\omega^{+}_{-} = \sqrt{(\omega_{0}^2 + g_{0})} \pm g_{0}$ with which we can naturally make the correspondence to the Rabi splitting energy: $\hbar\Omega = \hbar\omega_{+} - \hbar\omega_{-} = 2 g_{0}$

\section{Fitting of the Rabi-splitting Energies}

\begin{figure}[H]
\renewcommand{\thefigure}{S2}
\includegraphics[scale=0.7]{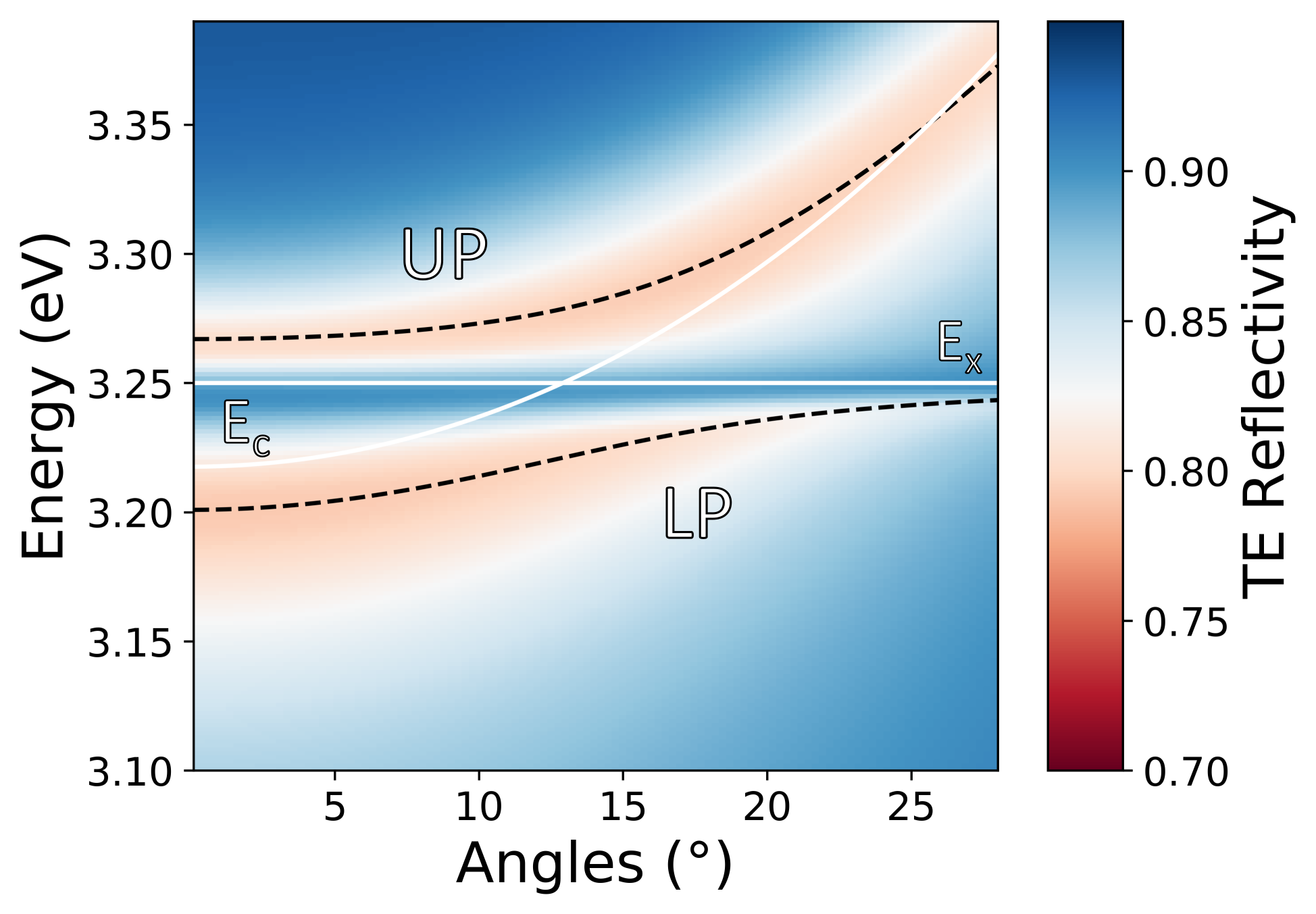}
\caption{\label{fig:FigureM1Fitting}  Simulated, angle-resolved, TE-polarized reflectivity map for a microcavity containing the material M1 as described in the main text. Overlaid are the fitting curves obtained using the analytical model in Ref 7. Overlaid solid white lines are the Excitons $E_{X_j}$ and cavity $E_{\rm C}$ modes, black dashed lines are the polaritons (LP/UP) obtained from fitting using the analytical model.}
\end{figure}

\begin{figure}[H]
\renewcommand{\thefigure}{S3}
\includegraphics[scale=0.7]{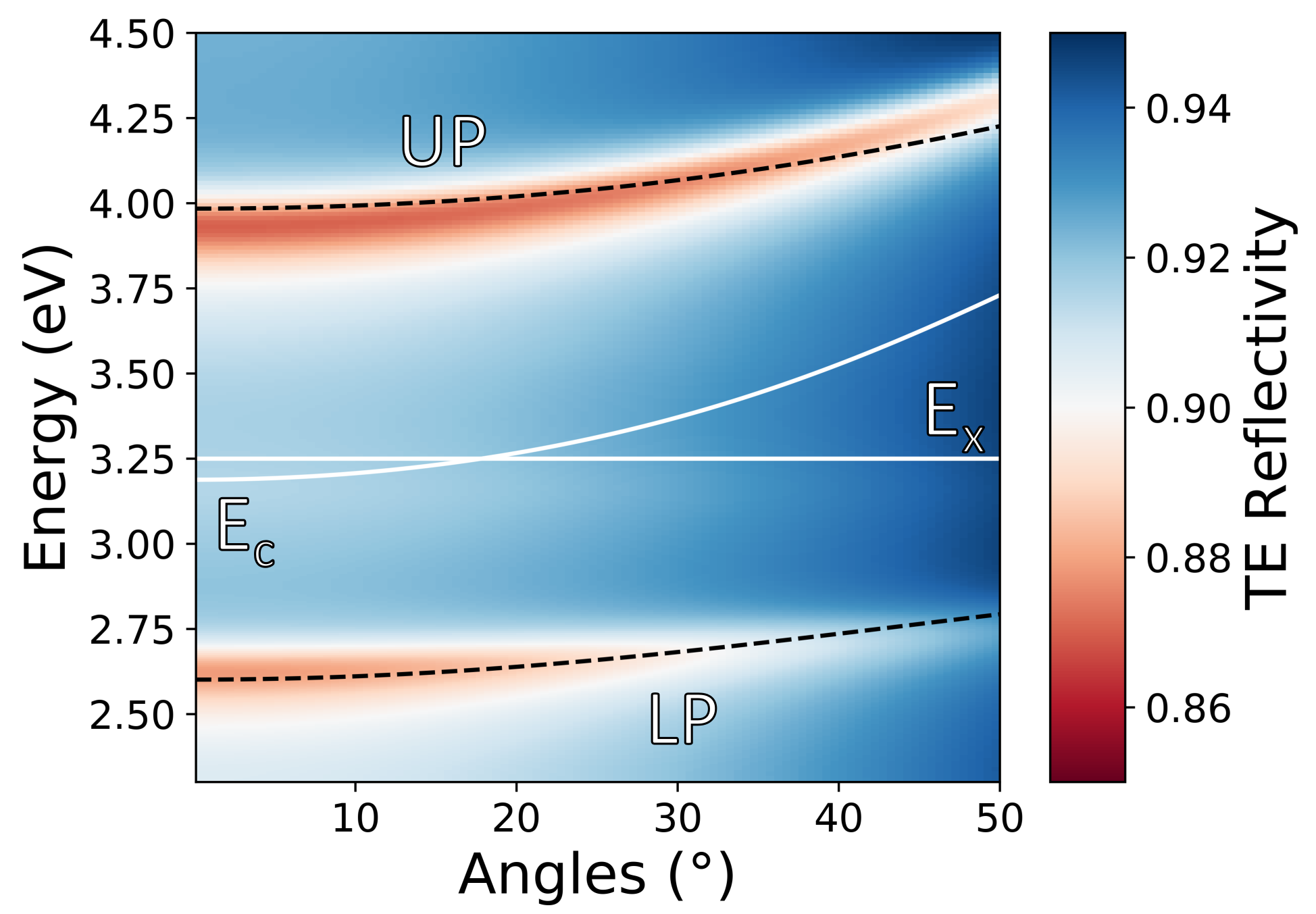}
\caption{\label{fig:FigureM2Fitting} Simulated, angle-resolved, TE-polarized reflectivity map for a microcavity containing the material M2 as described in the main text. Overlaid are the fitting curves obtained using the analytical model in Ref 7. Overlaid solid white lines are the Excitons $E_{X_j}$ and cavity $E_{\rm C}$ modes, black dashed lines are the polaritons (LP/UP) obtained from fitting using the analytical model.}
\end{figure}

All the Rabi-splitting energies $\hbar\Omega_{\rm R}$ mentioned in the text are derived using a Hopfield-Agranovich model\cite{Hopfield1958,Agranovich1957,Ciuti2005,LeRoux2019} presented in detail in Ref 7. The fitting results for the structure presented in Figure 2 (c) and Figure 3 (c) of the main text are respectively shown in Figure~\ref{fig:FigureM1Fitting} and Figure~\ref{fig:FigureM2Fitting}. The obtained $\hbar\Omega_{\rm R}$ values are respectively $\hbar\Omega_{{\rm R}_{M1}} = 58 \, \rm meV$ and $\hbar\Omega_{{\rm R}_{M2}} = 1.38 \, \rm eV$.

\section{Experimental Analysis}

\begin{figure}[H]
\renewcommand{\thefigure}{S4}
\includegraphics[scale=0.15]{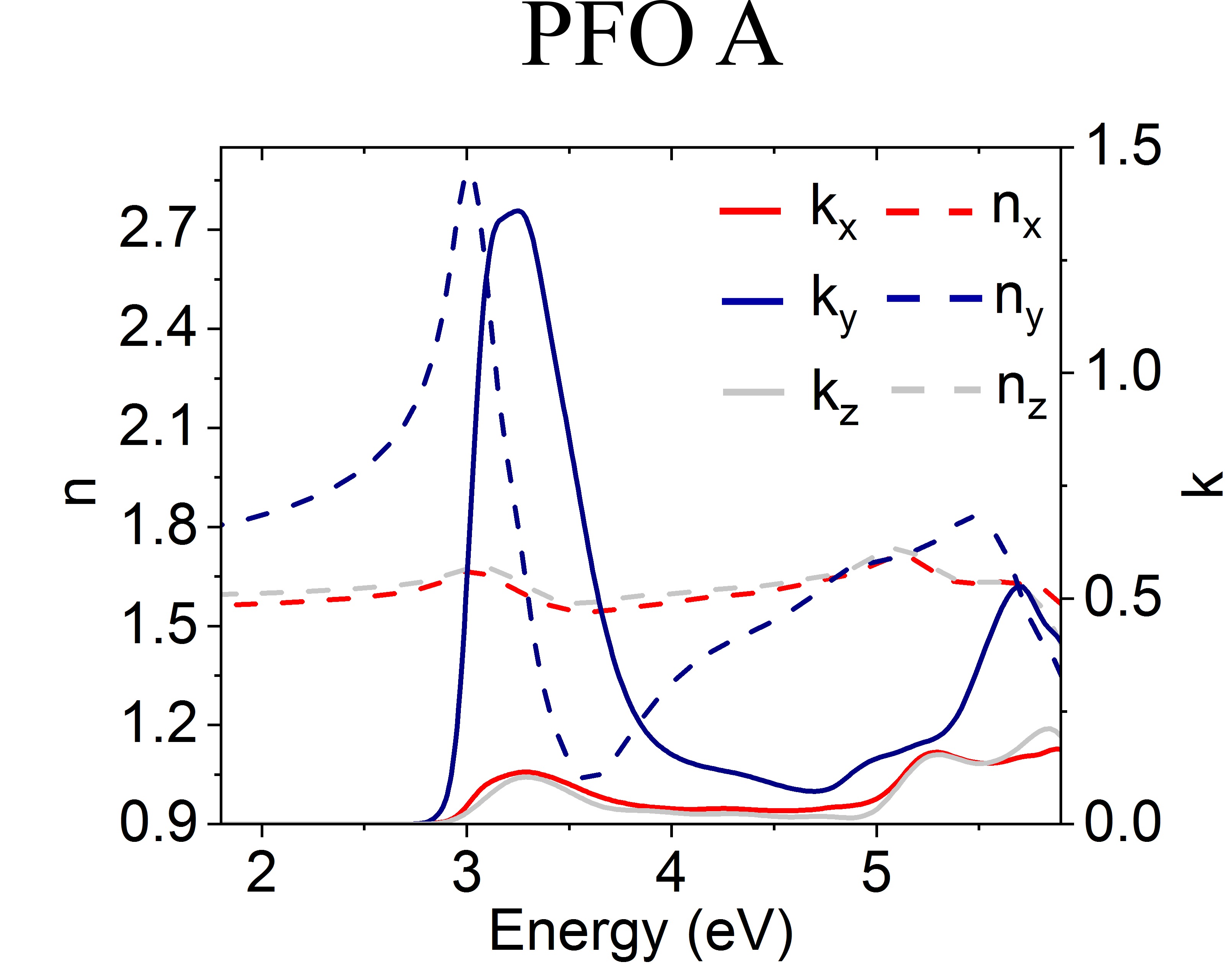}
\caption{\label{fig:FigurePFOIndices} (a) In red, blue and grey respectively ($n_ y$, $k_y$), ($n_x$, $k_x$) and ($n_z$, $k_z$)  optical components for aligned PFO reproduced from Ref 7. Dashed lines give the real component of the complex refractive index $\tilde{n} = n + {\rm i}k$, solid lines the imaginary component.}
\end{figure}

\begin{figure}[H]
\renewcommand{\thefigure}{S5}
\includegraphics[scale=0.7]{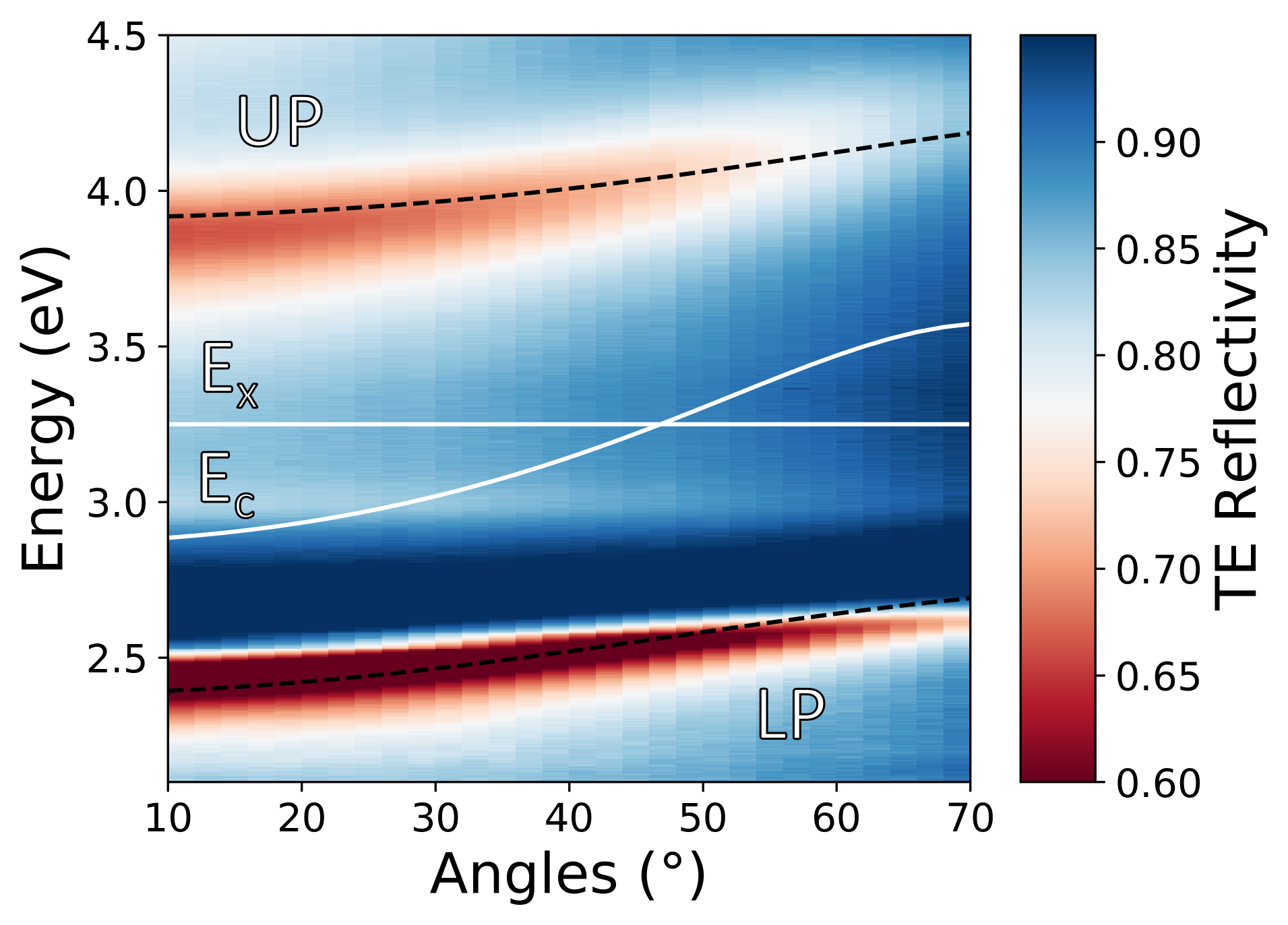}
\caption{\label{fig:FigurePFOExpFitting} Experimental, angle-resolved, TE-polarized reflectivity map for a microcavity containing aligned PFO as described in the main text. Overlaid are the fitting curves obtained using the analytical model in Ref 7. Overlaid solid white lines are the Excitons $E_{X_j}$ and cavity $E_{\rm C}$ modes, black dashed lines are the polaritons (LP/UP) obtained from fitting using the analytical model.}
\end{figure}

\begin{figure}[H]
\renewcommand{\thefigure}{S6}
\includegraphics[scale=0.7]{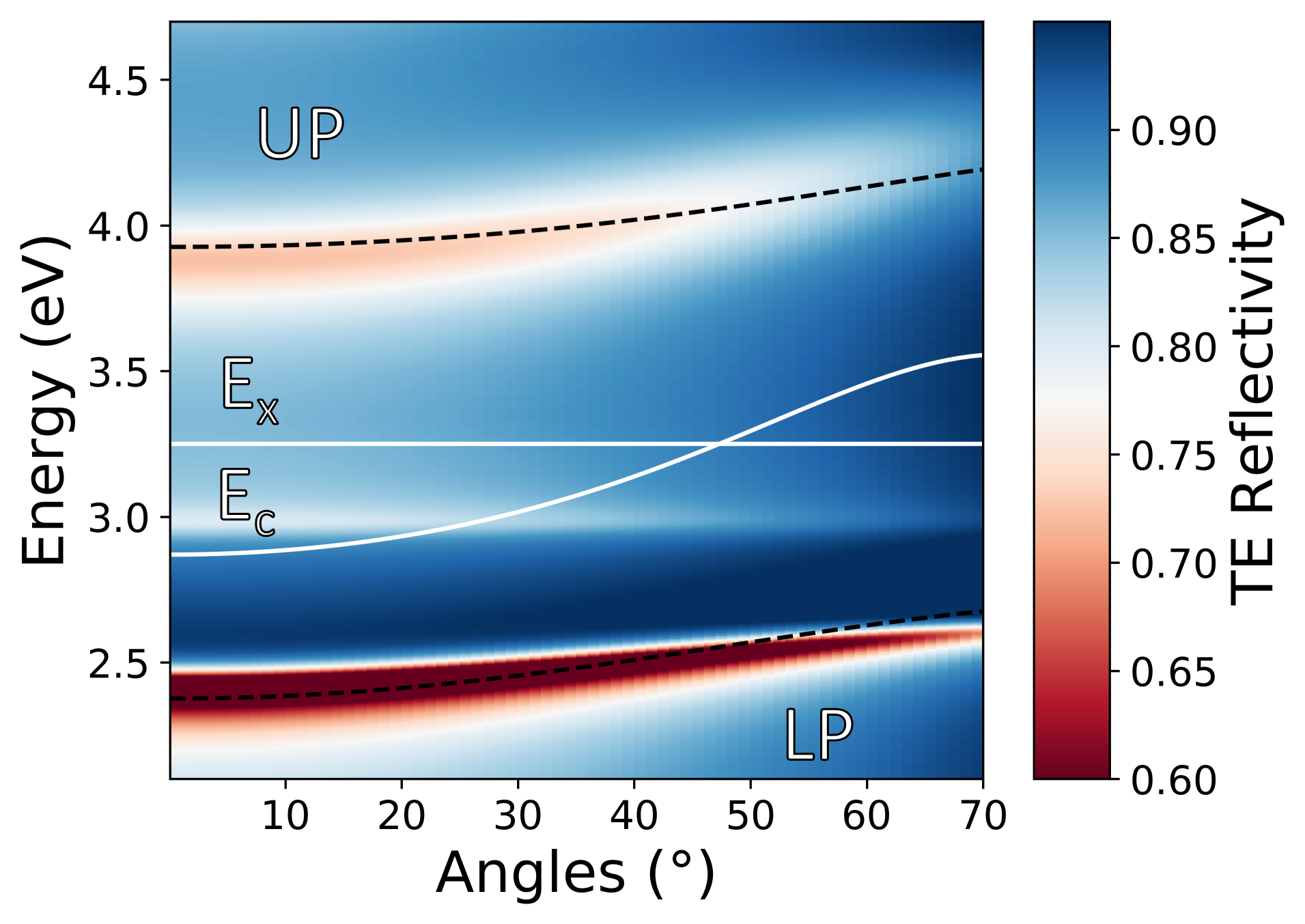}
\caption{\label{fig:FigurePFOSimFitting} Simulated, angle-resolved, TE-polarized reflectivity map for a microcavity containing aligned PFO as described in the main text. Overlaid are the fitting curves obtained using the analytical model in Ref 7. Overlaid solid white lines are the Excitons $E_{X_j}$ and cavity $E_{\rm C}$ modes, black dashed lines are the polaritons (LP/UP) obtained from fitting using the analytical model.}
\end{figure}

The fitting results for the structure presented in Figure 5 (a) of the main text are respectively shown in Figure~\ref{fig:FigurePFOExpFitting} (using the experimental values) and Figure~\ref{fig:FigurePFOSimFitting} (using simulated values). The obtained $\hbar\Omega_{\rm R}$ values are respectively $\hbar\Omega_{\rm R_{PFO_{exp}}} = 1.48 \, \rm eV$ and $\hbar\Omega_{\rm R_{PFO_{exp}}} = 1.50 \, \rm eV$.

\bibliography{manuscriptbiblioSI}